\documentclass[12pt]{article}
\title{Muon g-2 anomaly and extra interaction of composite 
Higgs in a dynamically broken electroweak theory}
\author{B.A. Arbuzov\\
{\it Skobeltsyn Institute for Nuclear Physics, Moscow State
University}\\ {\it 119899 Moscow, Russia}}
\date{}
\newcommand{\be}{\begin{equation}}
\newcommand{\ee}{\end{equation}}
\newcommand{\beq}{\begin{eqnarray}}
\newcommand{\eeq}{\end{eqnarray}}
\newcommand{\nn}{\nonumber}
\newcommand{\bi}{\bibitem}
\begin{document}
\maketitle
\begin{quote}
 In electroweak theory without elementary Higgs scalars 
existence of a solution, which breaks initial 
symmetry is shown.  A composite scalar doublet serves 
as a substitute for usual Higgs. The mass of the 
surviving Higgs scalar is predicted to be  
$M_H = 117 \pm 7\,GeV$. The resulting  theory 
contains realistic $W,\,Z,\,t$ masses and composite 
massive Higgs. Parameters used 
are $g$, $\theta_W$, cut-off $\Lambda\,\simeq\,3500\,GeV$ 
and  triple gauge constant $\lambda_V\,\simeq\,-0.034$. 
Stable nontrivial solution for extra interactions of 
the Higgs particle gives contribution to muon 
$g-2$: $\Delta a = 49.4\cdot 10^{-10}$ that agrees with 
the recently reported anomaly. The extra interaction leads 
to a number of experimental consequences for B.R. of Higgs 
decay and its production cross-sections in $e^+e^- $ and 
$\bar p p$ collisions including recommendation for looking 
for Higgs via decay $H \to 2\,\gamma$. 
\end{quote}
\newpage

\section{Introduction}

The widely popular Higgs mechanism~\cite{Higgs} of the 
electroweak symmetry breaking
needs initial scalar fields, which look rather less 
attractive, than the well-known gauge interactions of 
vector and spinor fields. Experimental facilities now 
approach 
the region of possible discovery of
the Higgs. There are even indications in favour of the Higgs 
mass $M_H\,=\,115^{+1.3}_{-0.9}\,GeV$~\cite{New}.
In view of these considerations it may be useful to
study once more possibilities, which differ from the standard 
Higgs mechanism and to look for experimental consequences of 
these possibilities. On the other hand there are serious 
indications on a discrepancy with theoretical values in 
measurements of the muon anomalous magnetic 
moment~\cite{muon}.

In the present work\footnote[1]{The work is supported in 
part by
grants "Universities of Russia" 015.02.02.05 (990588) and 
RFBR 01-02-16209.}
we consider a model, which might serve a substitute for the 
Standard Model in respect to symmetry breaking. The model 
leads 
to a number of effects, which differ from SM results, in 
particular, it gives reasonable effect for the muon $g-2$. 

We are basing here on the proposal,
which was done in works~\cite{Arb00},~\cite{Arb01}, the main 
point of which consists in application 
of N.N. Bogolyubov method of quasi-averages~\cite{Bog}
to a simple model. Let us summarize the essence of the
approach~\cite{Arb00} with some necessary corrections.
Namely we consider $U(1)$
massless gauge field $A_\mu$ and also massless spinor field 
$\psi$, which interact in the following way
\beq
& &L\,=\,\frac{\imath}{2}\bigl(\bar \psi \gamma_\rho 
\partial_\rho \psi\,-\,\partial_\rho \bar \psi \gamma_\rho 
\psi \bigr)\,-\,\frac{1}{4}A_{\mu \nu}A_{\mu \nu}\,+\nn\\
& &+\,e_{L}\bar \psi_L \gamma_\rho \psi_L A_\rho\,
+\,\,e_{R}\bar \psi_R \gamma_\rho \psi_R A_\rho\,;
\label{init}\\
& &A_{\mu\nu}\,=\,\partial_\mu A_\nu\,-\,\partial_\nu A_\mu
\,;\nn
\eeq
where as usually
$$
\psi_L\,=\,\frac{1+\gamma_5}{2}\,\psi\,;\qquad
\psi_R\,=\,\frac{1-\gamma_5}{2}\,\psi\,.
$$
We will not discuss here the problem of
the triangle axial anomaly, bearing in mind that in a
more realistic model one always can arrange cancellation of 
the anomalies by a suitable choice of fermions' charges, in 
the same way as it occurs in the Standard Model.

Now we start to apply Bogolyubov quasi-averages 
method~\cite{Bog}\footnote[2]{The quasi-averages method was 
first applied to quantum field theory problems in 
work~\cite{ATF}.}. In view of looking for symmetry breaking 
we add to~(\ref{init}) additional term
\be
\epsilon \cdot \bar \psi_L \psi_R\,\bar \psi_R \psi_L\,.
\label{eRL}
\ee

Now let us consider the theory with $\epsilon \neq 0$, 
calculate necessary quantities (averages) and only at this 
stage take limit $\epsilon \to 0$. In this limit, 
according~\cite{Bog}, we come to quasi-averages, which
not always coincide with the corresponding averages, which 
one obtains directly from the initial Lagrangian~(\ref{init}).

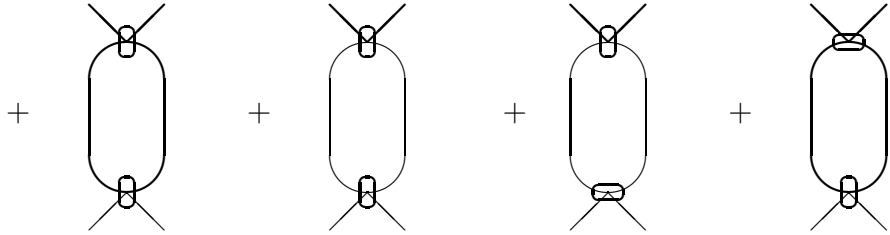
\begin{picture}(160,195)
{\thicklines
\put(7.5,192.5){\line(-1,1){5}}
\put(7.5,192.5){\line(1,1){5}}
\put(7.5,192.5){\oval(2,4)}}
\put(7.5,192.5){\line(-1,-1){5}}
\put(7.5,192.5){\line(1,-1){5}}
\put(16,192){=}
\put(24,192){$\Delta L_x\,;$}
{\thicklines
\put(50,192.5){\line(-1,1){5}}
\put(50,192.5){\line(1,1){5}}
\put(50,192.5){\oval(2,4)}
\put(50,192.5){\line(-1,-1){5}}
\put(50,192.5){\line(1,-1){5}}}
\put(60,192){=}
\put(68,192){$\Delta L_y\,;$}
\put(92.5,192.5){\line(-1,1){5}}
\put(92.5,192.5){\line(1,1){5}}
{\thicklines
\put(92.5,192.5){\oval(2,4)}}
\put(92.5,192.5){\line(-1,-1){5}}
\put(92.5,192.5){\line(1,-1){5}}
\put(102.5,192){=}
\put(110,192){$\Delta L_z\,.$}
{\thicklines
\put(7.5,172.5){\line(-1,1){5}}
\put(7.5,172.5){\line(1,1){5}}
\put(7.5,172.5){\oval(2,4)}}
\put(7.5,172.5){\line(-1,-1){5}}
\put(7.5,172.5){\line(1,-1){5}}
\put(18,172){=}
{\thicklines
\put(30,172.5){\line(-1,1){5}}
\put(30,172.5){\line(1,1){5}}
\put(30,172.5){\circle*{1}}}
\put(30,172.5){\line(-1,-1){5}}
\put(30,172.5){\line(1,-1){5}}
\put(29,163){$\epsilon$}
\put(40.5,172){+}
{\thicklines
\put(52.5,172.5){\line(-1,1){5}} \put(62.5,172.5)
{\oval(20,10)[t]}
\put(72.5,172.5){\line(1,1){5}}\put(52.5,172.5)
{\oval(2,4)}
\put(72.5,172.5){\oval(2,4)}}
\put(52.5,172.5){\line(-1,-1){5}} \put(72.5,172.5)
{\line(1,-1){5}}
\put(52.5,172.5){\line(1,0){20}}
\put(83,172){+}
{\thicklines
\put(95,172.5){\line(-1,1){5}} \put(105,172.5)
{\oval(20,10)[t]}
\put(115,172.5){\line(1,1){5}}\put(95,172.5){\oval(2,4)}
\put(115,172.5){\oval(2,4)}}
\put(95,172.5){\line(1,-1){5}} \put(115,172.5)
{\line(-1,-1){5}}
\put(95,172.5){\line(1,0){20}}
\put(125.5,172){+}
\put(2.5,142){+}
{\thicklines
\put(18.5,152.5){\line(-1,1){5}}
\put(18.5,152.5){\line(1,1){5}}
\put(18.5,142.5){\oval(10,20)}
\put(18.5,152.5){\oval(2,4)}
\put(18.5,132.5){\oval(2,4)}}
\put(18.5,132.5){\line(-1,-1){5}}
\put(18.5,132.5){\line(1,-1){5}}

\put(34.5,142){+}
{\thicklines
\put(50.5,152.5){\line(-1,1){5}}
\put(50.5,152.5){\line(1,1){5}}
\put(50.5,152.5){\oval(2,4)}
\put(50.5,132.5){\oval(2,4)}}
\put(50.5,132.5){\line(-1,-1){5}}
\put(50.5,132.5){\line(1,-1){5}}
\put(50.5,142.5){\oval(10,20)}
\put(68.5,142){+}
{\thicklines
\put(82.5,152.5){\line(-1,1){5}}
\put(82.5,152.5){\line(1,1){5}}
\put(82.5,152.5){\oval(2,4)}
\put(82.5,132.5){\oval(4,2)}}
\put(82.5,132.5){\line(-1,-1){5}}
\put(82.5,132.5){\line(1,-1){5}}
\put(82.5,142.5){\oval(10,20)}
\put(98.5,142){+}
{\thicklines
\put(114.5,152.5){\line(-1,1){5}}
\put(114.5,152.5){\line(1,1){5}}
\put(114.5,142.5){\oval(10,20)}
\put(114.5,152.5){\oval(4,2)}
\put(114.5,132.5){\oval(2,4)}}
\put(114.5,132.5){\line(-1,-1){5}}
\put(114.5,132.5){\line(1,-1){5}}
\put(75,117){(a)}
{\thicklines
\put(7.5,102.5){\line(-1,1){5}}
\put(7.5,102.5){\line(1,1){5}}
\put(7.5,102.5){\oval(2,4)}
\put(7.5,102.5){\line(-1,-1){5}}
\put(7.5,102.5){\line(1,-1){5}}}
\put(18,102){+}
{\thicklines
\put(30,102.5){\line(-1,1){5}}
\put(30,102.5){\line(1,1){5}}
\put(30,102.5){\oval(4,2)}
\put(30,102.5){\line(-1,-1){5}}
\put(30,102.5){\line(1,-1){5}}}
\put(40,102){=}
{\thicklines
\put(52.5,102.5){\line(-1,1){5}} \put(62.5,102.5)
{\oval(20,10)[t]}
\put(72.5,102.5){\line(1,1){5}}\put(52.5,102.5){\oval(2,4)}
\put(72.5,102.5){\oval(2,4)}
\put(52.5,102.5){\line(-1,-1){5}} \put(72.5,102.5)
{\line(1,-1){5}}
\put(52.5,102.5){\line(1,0){20}}}
\put(83,102){+}
{\thicklines
\put(95,102.5){\line(-1,1){5}} \put(105,102.5)
{\oval(20,10)[t]}
\put(115,102.5){\line(1,1){5}}\put(95,102.5){\oval(2,4)}
\put(115,102.5){\oval(2,4)}
\put(95,102.5){\line(1,-1){5}} \put(115,102.5){\line(-1,-1)
{5}}
\put(95,102.5){\line(1,0){20}}}
\put(125.5,102){+}
\put(2.5,72){+}
{\thicklines
\put(18.5,82.5){\line(-1,1){5}}
\put(18.5,82.5){\line(1,1){5}}
\put(18.5,72.5){\oval(10,20)}
\put(18.5,62.5){\line(-1,-1){5}}
\put(18.5,62.5){\line(1,-1){5}}
\put(18.5,82.5){\oval(2,4)}
\put(18.5,62.5){\oval(2,4)}}
\put(34.5,72){+}
{\thicklines
\put(50.5,82.5){\line(-1,1){5}}
\put(50.5,82.5){\line(1,1){5}}
\put(50.5,62.5){\line(-1,-1){5}}
\put(50.5,62.5){\line(1,-1){5}}}
\put(50.5,72.5){\oval(10,20)}
{\thicklines
\put(50.5,82.5){\oval(2,4)}
\put(50.5,62.5){\oval(2,4)}}
\put(68.5,72){+}
{\thicklines
\put(82.5,82.5){\line(-1,1){5}}
\put(82.5,82.5){\line(1,1){5}}
\put(82.5,62.5){\line(-1,-1){5}}
\put(82.5,62.5){\line(1,-1){5}}
\put(82.5,72.5){\oval(10,20)}
\put(82.5,82.5){\oval(2,4)}
\put(82.5,62.5){\oval(4,2)}}
\put(98.5,72){+}
{\thicklines
\put(114.5,82.5){\line(-1,1){5}}
\put(114.5,82.5){\line(1,1){5}}
\put(114.5,72.5){\oval(10,20)}
\put(114.5,62.5){\line(-1,-1){5}}
\put(114.5,62.5){\line(1,-1){5}}
\put(114.5,82.5){\oval(4,2)}
\put(114.5,62.5){\oval(2,4)}}
\put(128.5,72){+}
\put(2.5,32){+}
{\thicklines
\put(18.5,42.5){\line(-1,1){5}}
\put(18.5,42.5){\line(1,1){5}}
\put(18.5,22.5){\line(-1,-1){5}}
\put(18.5,22.5){\line(1,-1){5}}
\put(18.5,42.5){\line(0,-1){20}}
\put(18.5,32.5){\oval(10,20)[l]}
\put(18.5,42.5){\oval(4,2)}
\put(18.5,22.5){\oval(4,2)}}
\put(34.5,32){+}
{\thicklines
\put(50.5,42.5){\line(-1,1){5}}
\put(50.5,42.5){\line(1,-1){5}}
\put(50.5,22.5){\line(-1,-1){5}}
\put(50.5,22.5){\line(1,1){5}}
\put(50.5,42.5){\line(0,-1){20}}
\put(50.5,32.5){\oval(10,20)[l]}
\put(50.5,42.5){\oval(4,2)}
\put(50.5,22.5){\oval(4,2)}}
\put(66.5,32){+}
{\thicklines
\put(82.5,32.5){\line(-1,1){5}}
\put(82.5,32.5){\line(-1,-1){5}}
\put(102.5,32.5){\line(1,1){5}}
\put(102.5,32.5){\line(1,-1){5}}}
\put(92.5,32.5){\oval(20,10)}
{\thicklines
\put(82.5,32.5){\oval(4,2)}
\put(102.5,32.5){\oval(4,2)}}
\put(118.5,32.5){+}
\put(2.5,7){+}
{\thicklines
\put(18.5,7.5){\line(-1,1){5}}
\put(18.5,7.5){\line(-1,-1){5}}
\put(38.5,7.5){\line(1,1){5}}
\put(38.5,7.5){\line(1,-1){5}}
\put(28.5,7.5){\oval(20,10)}
\put(18.5,7.5){\oval(4,2)}
\put(38.5,7.5){\oval(4,2)}}
\put(54.5,7){+}
{\thicklines
\put(70.5,7.5){\line(-1,1){5}}
\put(70.5,7.5){\line(-1,-1){5}}
\put(90.5,7.5){\line(1,1){5}}
\put(90.5,7.5){\line(1,-1){5}}
\put(80.5,7.5){\oval(20,10)}
\put(70.5,7.5){\oval(2,4)}
\put(90.5,7.5){\oval(4,2)}}
\put(106.5,7){+}
{\thicklines
\put(122.5,7.5){\line(-1,1){5}}
\put(122.5,7.5){\line(-1,-1){5}}
\put(142.5,7.5){\line(1,1){5}}
\put(142.5,7.5){\line(1,-1){5}}
\put(132.5,7.5){\oval(20,10)}
\put(122.5,7.5){\oval(4,2)}
\put(142.5,7.5){\oval(2,4)}}
\put(72,-4){(b)}
\end{picture}
\bigskip

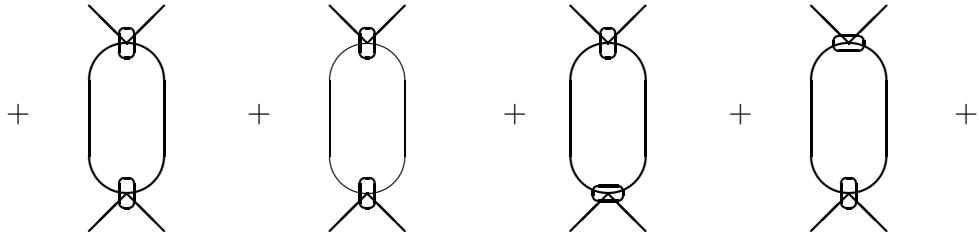
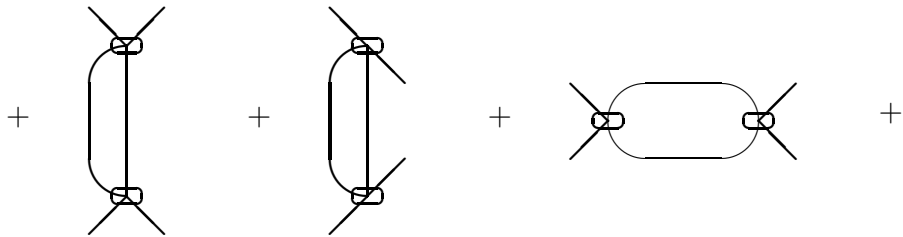
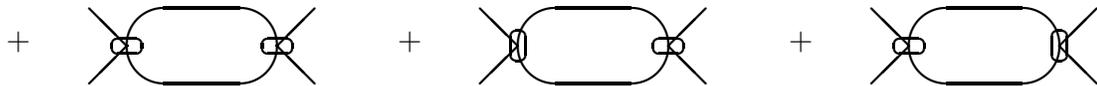
Figure 1. Diagrams for Eq.~(\ref{set}).
Thick lines represent left spinors $\psi_L$ and thin
lines represent right ones $\psi_R$; (a) corresponds to the 
first
equation of set~(\ref{set}) and (b) corresponds to the second
equation of~(\ref{set}). The third equation follows from (b)
by mutual exchange thick $\leftrightarrow$ thin

\newpage

Because of additional term~(\ref{eRL}) the following 
additional 
effective terms in Lagrangian~(\ref{init}) inevitably appear
\beq
& &\Delta L_x\,=\,x\cdot \bar \psi_L \gamma_\rho \psi_L\,\bar 
\psi_R \gamma_\rho \psi_R\,; \qquad \Delta L_y\,=\,\frac{y}{2}
\cdot \bar \psi_L \gamma_ 
\rho \psi_L\,\bar \psi_L \gamma_\rho \psi_L\,;\nn\\
& &\Delta L_z\,=\,\frac{z}{2}\cdot \bar \psi_R \gamma_\rho 
\psi_R\,  \bar \psi_R \gamma_\rho \psi_R\,.\label{psi}
\eeq
The corresponding vertices should have form-factors, which 
define effective cut-off $\Lambda$.
The origin of the cut-off is connected with self-consistent
solution of the corresponding dynamical equations. A 
realization of this mechanism will be shown below in 
Appendix. In the present work we just set upper limit at 
Euclidean $p^2\,=\,\Lambda^2$ in momentum integrals.

We consider compensation equations~\cite{Bog, Bog2}
(in other words, gap equations) for $x,\,y,\,z$ in one-loop
approximation (see Fig. 1). 
These equations are taken in the method of Fradkin 
full-vertices 
expansion~\cite{Fradkin}. The applicability of the method 
can be verified provided terms with more loops give 
much smaller contribution, than the first approximation 
does. We use here free lines, because one-loop terms give 
zero main contributions to fermion propagators. 
For the moment we neglect gauge 
field $A_\mu$ exchange loops, and obtain following set of 
equations
\beq
& & x\,=\,-\,\frac{\epsilon}{2}\,
+\,\frac{\Lambda^2}{16 \pi^2}\Bigl(\,-\, 3\,x^2\,-\,2\,x\,
(y + z)\Bigr)\,;\nn\\
& & y\,=\,\frac{\Lambda^2}{16 \pi^2}\Bigl( -\,x^2\,\Bigr)\,;
\qquad z\,=\,\frac{\Lambda^2}{16 \pi^2}\Bigl( -\,x^2\,\Bigr)
\,;\label{set}
\eeq
Let $X,\,Y,\,Z$ be dimensionless variables
\be
X\,=\,x\,\frac{\Lambda^2}{16 \pi^2}\,;\qquad
Y\,=\,y\,\frac{\Lambda^2}{16 \pi^2}\,;\qquad
Z\,=\,z\,\frac{\Lambda^2}{16 \pi^2}\,;\label{X}
\ee
Let us look for  solutions of set~(\ref{set}). At this stage 
we also set $\epsilon \to 0$. There is, of course, trivial 
solution $x\,=\,y\,=\,z\,=\,0\,$. In addition we have two 
nontrivial solutions.
\beq
& &Z_1\,=Y_1\,=\,-\,1\,;\qquad X_1\,=\,1\,;\label{1} \\
& &Z_2\,=Y_2\,=\,-\,\frac{1}{16}\,;\qquad X_2\,=\,-\,
\frac{1}{4}\,.\label{2}
\eeq

As we shall see further, just the second solution~(\ref{2})
will be the most interesting.

Now let us consider scalar bound states $(\bar \psi_L 
\psi_R,\,\bar \psi_R \psi_L)$.

\begin{picture}(160,30)
{\thicklines
\put(10,10){\line(-1,1){10}}
\put(10,10.5){\line(1,0){10}}}
\put(10,10){\circle*{3}}
\put(10,10){\line(-1,-1){10}}
\put(10,9.5){\line(1,0){10}}
\put(30,10){=}
{\thicklines
\put(50,10){\line(-1,1){10}}
\put(60,10){\oval(20,10)[t]}
\put(70,10.5){\line(1,0){10}}
\put(50,10){\oval(2,4)}}
\put(50,10){\line(-1,-1){10}}
\put(60,10){\oval(20,10)[b]}
\put(70,9.5){\line(1,0){10}}
\put(70,10){\circle*{3}}
\put(90,10){+}
{\thicklines
\put(120,10){\line(-1,1){13}}
\put(120,10.5){\line(1,0){10}}}
\put(120,10){\line(-1,-1){13}}
\put(120,9.5){\line(1,0){10}}
\put(120,10){\circle*{3}}
\multiput(110,0)(0,2){11}%
{\circle*{1}}

\end{picture}

\bigskip

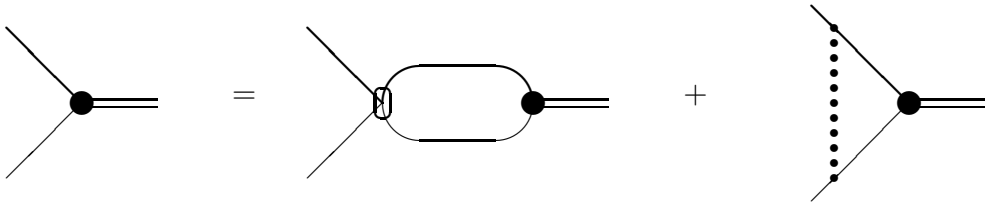
Figure 2. Diagram representation of Bethe-Salpeter equation 
for scalar
bound states. Dotted line represents gauge vector field.\\
\\
Without $e^2$ corrections 
we have from Bethe-Salpeter equation in one-loop 
approximation (see Fig. 2)
\beq
& &G\,=\,-\,4\,X\,F(\xi)\,G\,;\qquad \xi\,=\,\frac{k^2}
{4 \Lambda^2}\,;\qquad\mu\,=\,\frac{m^2}{\Lambda^2};
\label{eq0}\\
& &F(\xi)\,=\,1\, -\, \frac{5}{2}\,\xi\, +\, 2\xi\,
\log\frac{4 \xi}{1 + \xi}\,;\qquad \mu \ll \xi\,<\,1\,;\nn\\
& &F(\xi)\,=\,1\, +\, \frac{1}{3}\,\xi\, +\, 2\xi\,
\log \mu\,+\,O(\xi\mu,\,\xi^2)\,;\qquad \xi\leq \mu \,.\nn
\eeq
where $G\,=\,const$ is just the Bethe-Salpeter wave function. 
Here $m$ means a fermion mass, which will be shown to be 
nonzero, and $k^2$ is the scalar state Euclidean momentum 
squared, that is $k^2 > 0$ means tachyon mass of the scalar.
Function $F(\xi)$ decreases from the value $F(0) = 1$ 
with $\xi$ increasing. We see, that for solution~(\ref{2}) 
we have bound state with $k^2 = 0$ in full
correspondence with Bogolyubov-Goldstone theorem~\cite{Bog2},
\cite{Gold}. As for the first solution~(\ref{1}), there is no
solution of Eq.~(\ref{eq0}) at all.  So we have to  
concentrate our attention on the solution~(\ref{2}). Note, 
that there is an additional argument in favour of solution
~(\ref{2}). Namely, values $X$ and especially $Y,\,Z$ are 
small enough, so we may expect, that many-loop terms will 
not influence results strongly.

Now let us take into account vector boson corrections. 
Equation for the bound state~(\ref{eq0}) is modified due 
to two sources.  The first one corresponds to vector boson 
exchange loops in  set~(\ref{set}). Here corrections appear, 
for example, of such form
$$
\frac{6\,e_L\,e_R\,x}{16\,\pi^2}\,\log\frac{\Lambda^2}{p^2}\,.
$$
Here $p$ is the momentum of integration in
Eq.~(\ref{eq0}). In the present note we will restrict 
ourselves by main logarithmic approximation, in which such 
terms do not contribute to results. Indeed, due to simple 
relation
\be
\int_0^{\Lambda^2}\,x^n\,\log\frac{\Lambda^2}{x}\,dx\,=\,
\frac{(\Lambda^2)^{n+1}}{(n+1)^2}\,;\nn
\ee
there is no logarithms in final expressions. So the only 
important contribution consists in loop $e^2$
corrections to Eq.~(\ref{eq0}). In Landau gauge there
is only one such contribution: the triangle
diagram. We draw at Fig. 2 only this diagram, but the
gauge invariance of the result is checked by direct 
calculations in an arbitrary gauge.  So we now have
\be
G\,=\,F(\xi)\,G\,+\,
\frac{3\, e_L\, e_R}
{16\, \pi^2}\,\log \frac{\Lambda^2}{m^2}\,G\,.\label{eqe}
\ee
We see, that for small $e_L,\,e_R$ possible eigenvalues 
$\xi$ are also small. Then we have following condition for 
an eigenvalue
\be
F(\xi)\,+\,\frac{3\, e_L\, e_R}{16\, \pi^2}\,\log 
\frac{\Lambda^2}{m^2}\,=\,1\,;\label{eigen}
\ee
We see, that there is tachyon bound state in case $e^2$
contribution being positive.
Really, the eigenvalue condition for small $e_i^2$ reads 
\be
k^2\,=\,m_0^2\,=\,\frac{3\, e_L\, e_R}{8\, \pi^2}\,
\Lambda^2\,   
.\label{m0}
\ee
The result corresponds to main logarithmic approximation.
Thus provided $(e_L\, e_R) > 0$ we have scalar complex 
tachyon $\phi$ with negative mass squared $-\,m_0^2$.

We have the following vertices of interaction of $\phi$ 
with spinors
\be
G\,\Bigl(\bar \psi_R\,\psi_L\,\phi\,+
\,\bar \psi_L\,\psi_R\,\phi^*\Bigr)\,.\label{fint}
\ee
Normalization condition of Bethe-Salpeter equation gives
\be
G^2\,=\,\frac{16\,\pi^2}{\log\,(\Lambda^2/m^2)}\,.\label{c}   
\ee

Then we calculate box diagram with four scalar legs. This 
gives us effective constant $\lambda$, which enters into 
additional term
\be
\Delta\,L\,=\,-\,\lambda\,(\phi^*\,\phi)\,(\phi^*\,\phi)\,;
\qquad
\lambda\,=\,\frac{G^4}{16\,\pi^2}\,\log\,(\Lambda^2/m^2).
\label{lam}
\ee
Now we come to the usual Higgs model~\cite{Higgs} with 
$m_0^2$~(\ref{m0}), $\lambda$~(\ref{lam}) and $\phi$ charge 
$e_L\,-\,e_R$.

Thus from expressions~(\ref{m0}, \ref{lam}) we have usual
vacuum average of $\sqrt{2}\,Re\,\phi\,=\,\eta$
\be
\eta^2\,=\,\frac{m_0^2}{\lambda}\,=\,\frac{3\,e_L\,e_R}{128\,
\pi^4}\,\Lambda^2\,\log\,\frac{\Lambda^2}{m^2}\,.\label{eta}
\ee
   
The vector boson mass duly arises and it reads as follows
\be
M^2\,=\,\frac{3\,(e_L - e_R)^2\,e_L\,e_R}{128\,\pi^4}\,
\Lambda^2\,\log\,\frac{\Lambda^2}{m^2}\,.\label{M}
\ee
   
Interaction~(\ref{fint}) leads to spinor mass $m$
\be
m\,=\,\frac{G\,\eta}{\sqrt{2}}\,;\quad
m^2\,=\,\frac{3\,e_L\,e_R}{8\,\pi^2}\,\Lambda^2\,.\label{m}
\ee
   
Thus, we obtain the result, that initially massless model of 
interaction of a spinor with a vector becomes after the 
symmetry breaking just a close analog of the Higgs model. 
We have now vector boson mass~(\ref{M}), spinor mass
~(\ref{m}) and a scalar bound state with mass $\sqrt{2}\,m_0$,
\be
m^2_H\,=\,2\,m_0^2\,=\,\frac{3\, e_L\, e_R}{4\, \pi^2}\,
\Lambda^2\,.\label{mH}
\ee
   
We would formulate qualitative result of the simple model 
study as follows: in the massless model with Lagrangian
~(\ref{init}) with $(e_L \cdot e_R)\,>\,0$ there arises 
fermion-antifermion condensate, which defines masses 
$M,\,m,\,M_H$ according to~(\ref{M}, \ref{m}, \ref{mH}).   
   
Note, that variants of dynamical breaking of the electroweak 
symmetry without elementary scalars were considered in 
various 
aspects (see, e.g. paper~\cite{Arb92}). The possibility of
scalars being composed of fundamental spinors was considered 
e.g. in well-known paper~\cite{Ter}.

\section{Realistic model in the leading order of $1/N_c$ 
expansion}

Here we will proceed in the same line, but choose more 
realistic model, including one coloured left doublet $\psi_L\,
=\,(t_L,\,b_L)$ and two right singlets $t_R,\,b_R$, which 
simulate the heaviest quark pair. They interact with 
$SU(2)\times U(1)$ gauge bosons in standard way, so that the 
initial Lagrangian corresponds to the Standard Model with one 
heavy quark (initially massless) generation without Higgs 
sector 
and looks like
\beq
& &L\,=\,
\frac{\imath}{2}\bigl(\bar \psi_L \gamma_\rho 
\partial_\rho \psi_L\,
-\,\partial_\rho \bar \psi_L \gamma_\rho \psi_L \bigr)\,+\,
\frac{\imath}{2}\bigl(\bar t_R \gamma_\rho \partial_\rho 
t_R\,-
\,\partial_\rho \bar t_R \gamma_\rho t_R \bigr)\,+\nn\\
& &+\,\frac{\imath}{2}\bigl(\bar b_R \gamma_\rho \partial_
\rho b_R\,
-\,\partial_\rho \bar b_R \gamma_\rho b_R \bigr)\,-
\,\frac{1}{4}W_{\mu \nu}^a W_{\mu \nu}^a\,
-\,\frac{1}{4}B_{\mu \nu} B_{\mu \nu}\,-\nn\\
& &-\,\frac{g}{2}\,\bar \psi_L \tau^a\gamma_\rho \psi_L 
W^a_\rho\,
-\,\frac{g\tan \theta_W}{6 }\bar \psi_L \gamma_\rho 
\psi_L B_\rho\,
-\,\frac{2 g \tan \theta_W}{3}\bar t_R \gamma_\rho t_R 
B_\rho\,+\nn\\
& &+\,\frac{g \tan \theta_W}{3}\bar b_R \gamma_\rho b_R 
B_\rho\,;\qquad g =
\frac{e}{\sin \theta_W};\label{initr}\\
& &W_{\mu\nu}^a = \partial_\mu W_\nu^a - \partial_\nu 
W_\mu^a + g\,\epsilon^{abc}\,W^b_\mu\,W^c_\nu\,;\;
B_{\mu\nu} = \partial_\mu B_\nu - \partial_\nu B_\mu\,.\nn
\eeq
Notations here are usual ones and quarks $t$ and $b$
are colour triplets.

Let us look for spontaneous four-fermion interactions
\beq
& &x_1\,\bar \psi_L^\alpha \gamma_\rho \psi_{L \alpha}\,
\bar t_R^\beta \gamma_\rho t_{R \beta};\quad
x_2\,\bar \psi_L^\alpha \gamma_\rho \psi_{L \alpha}\,
\bar b_R^\beta \gamma_\rho b_{R \beta};\nn\\
& &\bar x_1\,\bar \psi_L^\alpha \gamma_\rho \psi_{L \beta}\,
\bar t_R^\beta \gamma_\rho t_{R \alpha};\quad
\bar x_2\,\bar \psi_L^\alpha \gamma_\rho \psi_{L \beta}\,
\bar b_R^\beta \gamma_\rho b_{R \alpha};\nn\\
& &\frac{y_1}{2}\,\bar \psi_L^{\alpha\,s} \gamma_\rho 
\psi_{L \alpha\,s}\,\bar \psi_L^{\beta\,r}\gamma_\rho 
\psi_{L \beta\,r}\,;\quad \frac{y_2}{2}\,\bar \psi_L^
{\alpha\,s} \gamma_\rho \psi_{L \alpha\,r}\,
\bar \psi_L^{\beta\,r} \gamma_\rho \psi_{L \beta\,s}\,;
\label{nset}\\
& &\frac{z_1}{2}\,\bar t_R^\beta \gamma_\rho t_{R \beta}\,
\bar t_R^\alpha \gamma_\rho t_{R \alpha}\,;\quad
\frac{z_2}{2}\,\bar b_R^\alpha \gamma_\rho b_{R \alpha}\,   
\bar b_R^\beta \gamma_\rho b_{R \beta}\,;\nn\\
& &z_{12}\,\bar t_R^\beta \gamma_\rho t_{R \beta}\,
\bar b_R^\alpha \gamma_\rho b_{R \alpha}\,;\quad
\bar z_{12}\,\bar t_R^\beta \gamma_\rho t_{R \alpha}\,
\bar b_R^\alpha \gamma_\rho b_{R \beta}\,;\nn\\
& &\bar y_1\,=\,y_2\,;\quad \bar y_2\,=\,y_1\,;\quad
\bar z_i\,=\,z_i\,.\nn
\eeq
Here $\alpha,\,\beta$ are colour indices and $s,\,r$ are weak 
isotopic indices. Symbol $\bar a$ means interchange of colour 
summation just in the way as it is done for $x_1,\,x_2$. In 
the same way, as earlier~(\ref{X}) we introduce dimensionless 
variables
\be
X_i\,=\,\frac{x_i\, \Lambda^2}{16\,\pi^2}\,;\quad Y_i\,=\,
\frac{y_i\, \Lambda^2}{16\,\pi^2}\,;\quad
Z_i\,=\,\frac{z_i\, \Lambda^2}{16\,\pi^2}\,.
\label{Xi}
\ee 
From diagram set of equations Fig. 1 we could obtain a 
complicated set of algebraic equations. Its form is 
marvelously simplified in the main order of 
$1/N_c$ expansion. So here we consider just this 
approximation. Thus for the beginning we get 
equations for $Y_1,\,Y_2$
\beq
& &Y_1\,=\,\frac{N_c}{2}\,\Bigl(-\,Y_2^2-2 Y^2_1-X^2_1-
X^2_2-2 Y_1\,Y_2\Bigr)\,;\nn\\
& &Y_2\,=\,\frac{N_c}{2}\,\Bigl(-\,Y_2^2-2 Y^2_1-X^2_1-
X^2_2-2 Y_1\,Y_2\Bigr)\,.\label{y}
\eeq
This means, that $Y_1\,=\,Y_2\,=\,Y$. For other variables 
we have
\beq
& &X_1\,=\,-\,N_c\,\Bigl(3 X_1\,Y+2 X_1 Z_1 + 
X_2 Z_{12}\Bigr)\,;\nn\\
& &X_2\,=\,-\,N_c\,\Bigl(3 X_2\,Y+2 X_2 Z_2 + 
X_1 Z_{12}\Bigr)\,;\nn\\
& &Y\,=\,-\,\frac{N_c}{2}\,\Bigl(5 Y^2+X_1^2+
X_2^2\Bigr)\,;\quad
\bar Y\,=\,Y;\;\bar Z_1\,=\,Z_1;\;\bar Z_2\,=\,Z_2;
\label{Nc}\\
& & Z_1\,=\,-\,\frac{N_c}{2}\,\Bigl(4\,Z_1^2 + 2\,X_1^2 
+ Z^2_{12}\Bigr);\quad
Z_2\,=\,-\,\frac{N_c}{2}\,\Bigl(4\,Z_2^2
+ 2\,X_2^2 +
Z^2_{12}\Bigr);\nn\\
& & Z_{12}\,=\,-\,2\,N_c \Bigl( Z_1\,Z_2 + Z_1\,Z_{12} 
+ Z_2\,Z_{12}\Bigr);\quad \bar Z_{12}\,=\,-\,N_c\,\bar 
Z_{12}^2\,.\nn\\
& &\bar X_1\,=\,-4\,N_c\,\bar X_1^2\,;\quad
\bar X_2\,=\,-4\,N_c\,\bar X_2^2\,;\label{nt}
\eeq
The most important are equations for $\bar X_{1,2}$, 
which we mark
separately. If we consider Bethe-Salpeter equations for
$\bar \psi_L t_R$ and $\bar \psi_L b_R$ scalar states again
in the main order of $1/N_c$ expansion, we have
respectively
\be
G_1= -\,4\,N_c\,\bar X_1\,F(\xi)\,G_1;\;
G_2= -\,4\,N_c\,\bar X_2\,F(\xi)\,G_2;\;
\xi\,=\,\frac{k^2}{4 \Lambda^2}.\label{eqn}
\ee  
Here $F(\xi)$ is the same function as earlier~(\ref{eq0}).
This means that in case of a non-trivial solution of one of 
equations~(\ref{nt}) we  have zero mass Bogolyubov-Goldstone 
scalar doublet. Interaction with gauge vector  fields 
according to~(\ref{initr}) shifts
the level to positive or negative values of the mass squared.
Because of the bound state consisting of left and right 
spinors,
only interactions of $B$-boson enter to diagram Fig. 2.
So if $\bar X_1\,=\,-\,1/4 N_c$, we  have tachyon state with
\be
m_0^2\,=\,\frac{g^2 \tan^2\theta_W}{24\,\pi^2}\,\Lambda^2\,.
\label{m0n}
\ee
For other possibility, i.e. $\bar X_2\,=\,-\,1/4 N_c$ 
($\bar \psi_L\, b_R$ bound state), we have normal sign 
mass squared
$$
m_{L b}^2\,=\,\frac{g^2 \tan^2\theta_W}{48\,\pi^2}\,
\Lambda^2\,.
$$
As we already know, the first possibility leads to Higgs-like 
symmetry breaking, that is to negative minimum of the 
effective 
potential. As for the second case, there is no minima, so 
here 
we should take trivial solution $\bar X_2\,=\,0$. It is very 
important point. Really, we here obtain the explanation of 
why 
the $t$-quark is heavy and the $b$-quark is light. Only the 
first possibility corresponds to tachyon and so gives 
condensate, leading to creation of masses including
$t$-quark mass. The fact is connected with signs of 
interaction 
terms in~(\ref{initr}). For us here only $\bar X_1$ is 
important. 
It is easily seen from set~(\ref{Nc}) that for all other 
variables we can take trivial solution $X_1 = X_2 = ... = 0$. 

Next step is normalization of the bound state. Again 
following 
previous considerations~(\ref{fint}, \ref{c}) we have the 
following normalization condition
\be
\frac{N_c\,G^2_1}{16\,\pi^2}\,\log\frac{\Lambda^2}{m^2}\,=
\,1\,;
\label{norm}
\ee
Now we obtain fourfold scalar interaction constant $\lambda$
\be
\lambda\,=\,\frac{N_c\,G^4_1}{16\,\pi^2}\,\log\frac{\Lambda^2}
{m^2}\,.\label{lamn}
\ee

For the moment we have everything for effective Higgs 
mechanism 
in our variant of the Standard Model. We immediately come to 
classical scalar field density $\eta$
\be
\eta^2\,=\,\frac{m_0^2}{\lambda}\,;\label{etan}
\ee
where parameters are uniquely defined~(\ref{m0n}, \ref{norm}, 
\ref{lamn}). We also easily see, that constants of gauge 
interactions of the scalar doublet are just $g$ for 
interaction 
with $W$ and $g'\,=\,g\,\tan \theta_W$ for interaction with 
$B$ 
as usually. Now we have at once $W,\,Z$ and $t,\,b$-quark 
masses
\be
M_W\,=\,\frac{g\,\eta}{2}\,;\quad M_Z\,=\,\frac{M_W}
{\cos\theta_W}\,;\quad m_t\,=\,\frac{G_1\,\eta}{\sqrt{2}}\,;
\quad m_b\,=\,0\,.\label{MWt}
\ee
Mass of surviving Higgs scalar
\be
M_H^2\,=\,2\,m_0^2\,=\,2\,\lambda\,\eta^2\,=\,
\frac{N_c\,G_1^2\,m_t^2}{4\,\pi^2}\,\log\frac{\Lambda^2}
{m_t^2}\,.\label{Higgsn}
\ee
From~(\ref{m0n}) we have
\be
\Lambda^2\,=\,\frac{3\,\pi \cos^2\theta_W}{\alpha}\,M_H^2\,;
\label{Lambda}
\ee
where as usually $\alpha$ is the fine structure constant.
From here and from~(\ref{Higgsn}) we have
\be
R\,=\,\frac{3\,G_1^2}{4\,\pi^2}\,\biggl(\log R\,+\,
\log\frac{3\,\pi\,\cos^2\theta_W}{\alpha}\biggr)\,;
\quad R\,=\,\frac{M_H^2}{m_t^2}\,.\label{!}
\ee
  
Now let us discuss phenomenological aspects. We know all 
masses 
but that of the Higgs particle. To obtain it let us proceed 
as follows. From~(\ref{MWt}) we have
\be
G_1\,=\,\frac{g\,m_t}{\sqrt{2}\,M_W}\,;\quad g\,=\,\sqrt{\frac
{4\,\pi\,\alpha}{\sin^2\theta_W}}\,=\,0.648;\quad 
\alpha(M_Z)\,
=\,\frac{1}{129}\,.\label{G11}
\ee
Then for values $m_t\,=\,174 \pm 5\,GeV$, $M_W\,=\,80.3\,GeV$ 
we have $G_1\,=\,0.993 \pm 0.032$. Substituting this 
into~(\ref{!}) we have from solution of the equation
\be
M_H\,=\,117.1\, \pm\, 7.4\,GeV\,.\label{hor}
\ee
For example, for $m_t\,=\,173\,GeV$ we obtain 
$M_H\,=\,115.6\,GeV$. Result~(\ref{hor}) support indications 
of $115\,GeV$ Higgs~\cite{New}.

However value $G_1\,\simeq\, 1$ does not fit normalization 
condition~(\ref{norm}). From~(\ref{norm}) $G_1^2$ is rather 
about $8$ than about $1$. So we conclude, that condition
~(\ref{norm}) contradicts to the observed $t$-quark mass. We 
would emphasize , that the normalization condition is the 
only 
flexible point in our approach. The next section will be 
devoted to discussion of this problem.
   
\section{Normalization condition}

Scalar bound state in our approach consists of anti-doublet 
$(t_L,\,b_L)$ and singlet $t_R$. It evidently has weak 
isotopic spin $1/2$. Let us denote its field as $\phi$
\be
\phi\,=\,\Bigl(\bar b_L\,t_R \quad\bar t_L\,t_R \Bigr) \,;
\qquad \phi^+\,=\,  \Bigl(  \bar t_R\, b_L\quad\bar t_R\, 
t_L\Bigr)\,.\label{f}
\ee
From  the previous results we have the following terms
in the effective Lagrangian
\beq
& &N\,\frac{\partial \phi^+}{\partial x^\mu}\,
\frac{\partial \phi}{\partial x^\mu}\,+\, m_0^2\,\phi^+ 
\phi \,-\,\Lambda\,(\phi^+ \phi)^2 \,;\label{N}\\
& &N\,=\,\frac{N_c\,G^2_1}{16\,\pi^2}\,\log\frac{\Lambda^2}
{m^2}\,;\nn
\eeq
where parameters are defined in~(\ref{m0n}, \ref{lamn}). As 
for normalization parameter $N$, it does not satisfy us.
So we may study a possibility of quasi-averages effect
for $\phi$ interactions, which can change a coefficient 
afore the kinetic term. So, let us consider the following
possible interactions in addition to~(\ref{N})
\beq
& &L_{int}(W)\,=\,\frac{\xi_{01}}{2}\,\frac{\partial \phi^+}
{\partial x^\mu}\,\frac{\partial \phi}{\partial x^\nu}\,W^a_
{\mu \rho}\,W^a_{\nu \rho}\,+\,\frac{\eta_{01}}{2}\,\frac
{\partial \phi^+}{\partial x^\mu}\,\frac{\partial \phi}
{\partial x^\mu}\,W^a_{\nu \rho}\,W^a_{\nu \rho}\,;
\label{ffWW}\\
& &L_{int}(B)\,=\,\frac{\xi_{00}}{2}\,\frac{\partial \phi^+}
{\partial x^\mu}\,\frac{\partial \phi}{\partial x^\nu}\,
B_{\mu \rho}\,B_{\nu \rho}\,+\,\frac{\eta_{00}}{2}\,\frac
{\partial \phi^+}{\partial x^\mu}\,\frac{\partial \phi}
{\partial x^\mu}\,B_{\nu \rho}\,B_{\nu \rho}\,.\label{ffBB}
\eeq
In obtaining compensation equations we take into account 
possibility
of triple $W$ gauge bosons coupling, which was discussed in 
various
papers (e.g.~\cite{Hag}) and, in particular, in a variant 
of dynamical breaking
of the electroweak symmetry, which the author considered 
some time ago~\cite{Arb92}.
 
\begin{picture}(160,40)
\multiput(30,10)(0,2){7}%
{\circle*{1}}
\multiput(30,10)(-1.72,-1){7}%
{\circle*{1}}
\multiput(30,10)(1.72,-1){7}%
{\circle*{1}}
\put(30,10){\circle*{3}}
\put(50,10){=}
\multiput(80,20)(0,2){5}%
{\circle*{1}}
\multiput(70,2.8)(2,0){10}%
{\circle*{1}}
\multiput(80,20)(-1,-1.72){16}%
{\circle*{1}}
\multiput(80,20)(1,-1.72){16}%
{\circle*{1}}
\put(80,20){\circle*{3}}
\put(70,2.8){\circle*{3}}
\put(90,2.8){\circle*{3}}
\end{picture}
\bigskip

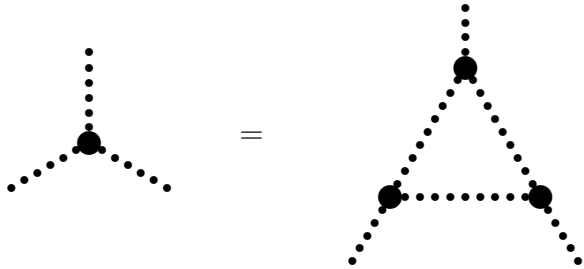
Figure 3. Diagram equation for triple gauge constant 
$\lambda_V$.\\
\\
In this variant the Bogolyubov quasi-averages 
method is
applied to possible origin of the triple interaction
\be
\frac{g \lambda_V}{M_W^2}\,\epsilon_{a b c}\,W_{\mu \nu}^a\,
W_{\nu \rho}^b\,
W_{\rho \mu}^c\,;\label{WWW}
\ee 
and corresponding one-loop compensation equation looks like 
(see Fig. 3)
\be
\lambda_V\,=\,\lambda_V\,\Biggl(\frac{g \lambda_V\,}{M_W^2}
\Biggr)^2\,
\frac{\Lambda^4}{128\,\pi^2}\,;\label{compW}
\ee
We have solutions of this equation: trivial one $\lambda_V\,
=\,0$ and two non-trivial ones
\be
\lambda_V\,=\,\pm\,\lambda_0\,;\quad
\lambda_0\,=\,\frac{8\,\sqrt{2}\, M_W^2}{g\,\Lambda^2}\,.
\label{lV}
\ee
In the following we assume, that genuine value of $\lambda_V$
 may differ from value~(\ref{lV}) in the range of 15-20\%.

Set of equations in one-loop approximation according to 
diagrams
presented at Fig. 4 looks like
\beq
& &\xi_1\,=\,-\,\frac{1}{3}\,\xi_1^2\,-\, \frac{1}{3}\,
\xi_1\,\eta_1\,-
\, \frac{1}{12}\,\eta_1^2\,
+\,\frac{4}{3}\, \zeta\,\xi_1 \,;\nn\\
& &\eta_1\,=\,-\,\frac{1}{48}\,\xi_1^2\,-\, \frac{1}{12}\,
\xi_1\,\eta_1\,-\,
\frac{1}{12}\,\eta_1^2\,+\,4\,\zeta\,\biggl( \frac{5}{6}\,
\xi_1\,+\,2\,
\eta_1 \biggr)\,;\label{xi1}\\
& &\xi_0\,=\,-\,\frac{1}{3}\,\xi_0^2\,-\, \frac{1}{3}\,
\xi_0\,\eta_0\,-
\, \frac{1}{12}\,\eta_0^2\,;\label{xi0}\\
& &\eta_0\,=\,-\,\frac{1}{48}\,\xi_0^2\,-\, \frac{1}{12}\,
\xi_0\,\eta_0\,-\,
\frac{1}{12}\,\eta_0^2\,;\nn\\
& &\xi_1\,=\,\frac{\Lambda^4}{16\,\pi^2}\,\xi_{01}\,;\quad
\eta_1\,=\,\frac{\Lambda^4}{16\,\pi^2}\,\eta_{01}\,;\quad
\zeta\,=\,\biggl(\frac{\lambda_V}{\lambda_0}\biggr)^2\,.\nn\\
& &\xi_0\,=\,\frac{\Lambda^4}{16\,\pi^2}\,\xi_{00}\,;\quad
\eta_0\,=\,\frac{\Lambda^4}{16\,\pi^2}\,\eta_{00}\,.\nn
\eeq
For parameters describing interaction of vector singlet $B$ 
there is   
no contribution of the triple vertex~(\ref{WWW}), because 
$B$ is
an Abelian field.

\begin{picture}(160,90)
{\thicklines
\put(0,70.5){\line(1,0){20}}}
\put(0,69.5){\line(1,0){20}}
\multiput(10,70)(-1.42,1.42){7}%
{\circle*{1}}
\multiput(10,70)(1.42,1.42){7}%
{\circle*{1}}
\put(10,70){\circle{3}}
\put(30,70){=}
{\thicklines
\put(40,70.5){\line(1,0){40}}}
\put(40,69.5){\line(1,0){40}}
\multiput(41.42,78.58)(1.42,-1.42){14}%
{\circle*{1}}
\multiput(78.58,78.58)(-1.42,-1.42){14}%
{\circle*{1}}
\put(50,70){\circle{3}}
\put(70,70){\circle{3}}
\put(90,70){+}
{\thicklines
\put(100,70.5){\line(1,0){40}}}
\put(100,69.5){\line(1,0){40}}
\multiput(110,70)(1.42,-1.42){8}%
{\circle*{1}}
\multiput(130,70)(-1.42,-1.42){8}%
{\circle*{1}}
\multiput(110,70)(1.42,1.42){6}%
{\circle*{1}}
\multiput(130,70)(-1.42,1.42){6}%
{\circle*{1}}
\put(110,70){\circle{3}}
\put(130,70){\circle{3}}
\put(145,70){+}
{\thicklines
\put(0,0.5){\line(1,0){40}}}
\put(0,-0.5){\line(1,0){40}}
\put(20,0){\circle{3}}
\multiput(20,0)(-1,1.72){16}%
{\circle*{1}}
\multiput(20,0)(1,1.72){16}%
{\circle*{1}}
\multiput(10,17.2)(2,0){11}%
{\circle*{1}}
\put(10,17.2){\circle*{3}}
\put(30,17.2){\circle*{3}}
\put(50,10){+}
{\thicklines
\put(60,0.5){\line(1,0){40}}}
\put(60,-0.5){\line(1,0){40}}
\put(80,0){\circle{3}}
\multiput(80,0)(-1,1.72){11}%
{\circle*{1}}
\multiput(80,0)(1,1.72){11}%
{\circle*{1}}
\multiput(70,17.2)(2,0){11}%
{\circle*{1}}
\put(70,17.2){\circle*{3}}
\put(90,17.2){\circle*{3}}
\multiput(70,17.2)(1.42,1.42){6}%
{\circle*{1}}
\multiput(90,17.2)(-1.42,1.42){6}%
{\circle*{1}}
\end{picture}
\bigskip

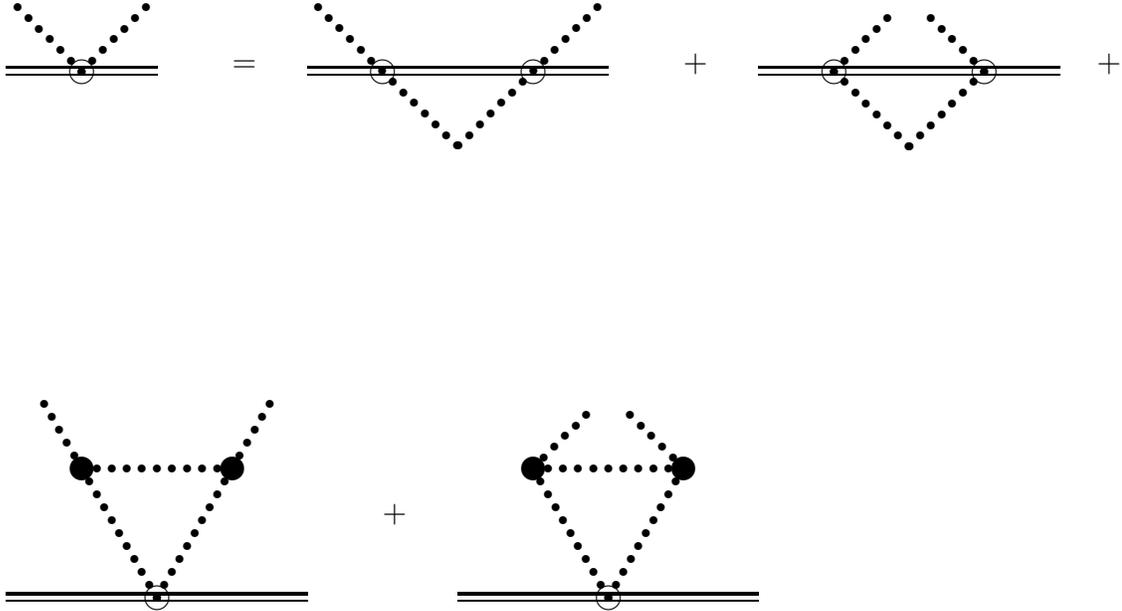
Figure 4. Diagrams representing equations for $\phi^+\phi 
W W$ vertices.\\
\\
\begin{picture}(160,40)
{\thicklines
\put(60,10.5){\line(1,0){40}}}
\put(60,9.5){\line(1,0){40}}
\put(80,10){\circle{3}}
\multiput(80,10)(-1,1.72){11}%
{\circle*{1}}
\multiput(80,10)(1,1.72){11}%
{\circle*{1}}
\multiput(70,27.2)(1.42,1.42){8}%
{\circle*{1}}
\multiput(90,27.2)(-1.42,1.42){7}%
{\circle*{1}}
\end{picture}
\bigskip

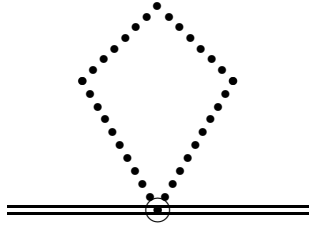
Figure 5. Diagram for contribution of $\phi^+\phi W W$ 
vertices to the
normalization condition.\\
\\
Parameters $\xi_i,\,\eta_i$ enter into normalization 
condition according to
account of diagram Fig 5. Now this condition takes the form
\be
\frac{N_c\,G^2_1}{16\,\pi^2}\,\log\frac{\Lambda^2}{m^2}\,+\,
\frac{9}{4}\,\Bigl(\xi_1\,+\,2\,\eta_1\Bigr)\,+\,\frac{3}{4}\,
\Bigl(\xi_0\,+\,2\,\eta_0\Bigr)\,=\,1\,;\label{normn}
\ee
From this expression we obtain
\be
G^2_1\,=\,\frac{(4 - 9 (\xi_1 +2 \eta_1) - 3 
(\xi_0 +2 \eta_0))\,4\,\pi^2}
{3\,\Bigl(\log (3\, \pi\,\cos^2\theta_W\,M_H^2)\,-\,
\log (\alpha\,m_t^2)\Bigr)}\,.\label{Gc}
\ee
We take $M_H$~(\ref{hor}) and obtain results, which are
presented at Table 1. 
\begin{center}
Table 1.\\
 Yukawa coupling $G_1$ and the $t$-quark mass\\ 
in dependence on $\lambda_V$.

\bigskip

\begin{tabular}{|l|c|c|l|c|l|c|}\hline
$\zeta$ & $\xi_1$ & $\eta_1$ & $G_1$ & $\lambda_V$ & $m_t\,
GeV$& 
$G_1(\xi_0,\eta_0 \ne 0)$\\ \hline
1 & 1.723 & - 0.820 & 2.67 & - 0.0286 &  467.8 & 5.31\\ \hline
1.2 & 3.056 & - 1.421 & 2.13  & - 0.0313 & 373.2 & 5.31 \\ 
\hline
1.4 & 4.371 & - 1.999 & 1.195 & - 0.0338 & 209.4 & 4.75\\ 
\hline
1.42 & 4.502 & - 2.056 & 1.047 & - 0.0341 & 183.4 & 4.71 \\ 
\hline
1.422 & 4.515 & - 2.062 & 1.031 & - 0.0341 & 180.6 & 4.71\\ 
\hline
1.424 & 4.528 & - 2.068 & 1.015 & - 0.0341 & 177.8 & 4.70\\ 
\hline
1.426 & 4.541 & - 2.073 & 0.998 & - 0.0342 & 174.8 & 4.70\\ 
\hline
1.428 & 4.554 & - 2.079 & 0.982 & - 0.0342 & 172.0 & 4.70\\ 
\hline
1.43 & 4.567 & - 2.085 & 0.964 & - 0.0342 & 168.9 & 4.69\\ 
\hline
0 & - 2.781 & - 0.215 & 8.392 & 0 & 1470.7 &\\ \hline
0 & 0 & 0 & 2.927 & 0 & 512.8 & \\ \hline
\end{tabular}
\end{center}
\bigskip
Table 1 contains results of calculations
for the case of trivial solution of set~(\ref{xi0}), i.e 
$\xi_0\,=
\,\eta_0\,=\,0$ (all columns but the last); this last column 
presents
results for nontrivial solution of
set~(\ref{xi0}): $\xi_0\,=\,-\,2.781,\,\eta_0\,=\,-\,0.215$.

What are criteria for choosing a solution?
The main criterion is provided by an energy density of a 
vacuum. We know, that
in case of appearance of scalar Higgs condensate $\eta$ the 
vacuum energy density reads
\be
E\,=\,-\,\frac{m_0^4}{4\,\lambda}\,;\label{E}
\ee
where parameters are defined in~(\ref{m0n}), (\ref{lamn}).
We see, that $\lambda$ is proportional to $G_1^4$, thus the 
smaller is $G_1$ and consequently $m_t$, the deeper becomes
the energy density minimum. Therefore we have to choose the
solution, which leads to  the minimal value of $m_t$.

From Table 1 we see, that for calculated value $\lambda_V = 
- \lambda_0$
we have minimal $m_t$ in comparison with two last lines. 
The last line
but one corresponds to non-trivial
solution with $\lambda_V\,=\,0$ and the last line give 
results for
completely trivial solution ($\Lambda\,=\,\xi\,=\,\eta\,=\,
0$). Both
ones give larger values for $m_t$ and thus are not suitable. 
The last column of the table also corresponds to non-minimal 
values of $G_1$
and thus the corresponding solutions are unstable.

However, $m_t \simeq 470\,GeV$ is too large, so let us look 
for values of
$\zeta = (\lambda_V/\lambda_0)^2$, which give realistic 
values of the $t$-quark mass. Namely,
values $\zeta \simeq 1.42 \div 1.43$,
($\lambda_V\,=\,-\,0.0342$) give
$G_1$ to be just in correspondence with relation~(\ref{G11}).
So, contradiction of the normalization condition with value 
of the
$t$-quark mass is removed. Thus we come to the conclusion, 
that we may  
not take into account of the normalization condition in 
considering
estimates of composite Higgs mass, as it is done in the 
previous section.

How we can interpret the situation, when the desirable value 
for
$\lambda_V$ is about 20\% larger, than the calculated one? 
We would state,
that corrections to the leading approximation, which is used 
here, might
be just of this order of magnitude. To estimate the effect 
of correction
let us consider contribution of usual electroweak triple 
$W$ vertex
$$
V_{\mu\nu\rho}(p,q,k)\,=\,g\,\Bigl(g_{\mu \nu}(q_\rho - 
p_\rho)\,+
\,g_{\nu\rho}(k_\mu-q_\mu)\,+\,g_{\rho\mu}(p_\nu-k_\nu)\Bigr);
$$
(where $p,\,q,\,k$ and $\mu,\,\nu,\,\rho$ are as usually 
incoming momenta
and indices of $W$s) to compensation equation~(\ref{compW}), 
which defines
$\lambda_V$. In addition to diagram Fig. 3 we take three 
diagrams, in which
one of vertices is changed to $V_{\mu\nu\rho}$. We have to 
add also
contribution of four-$W$ vertex, which is contained in 
triple vertex~(\ref{WWW}). This contribution restores 
the gauge invariance of the result.
Thus we obtain a contribution, which is proportional to 
$g/\pi$.
Namely, instead of~(\ref{compW}) we now have
$$
\lambda_V\,=\,\lambda_V\,\Biggl(\Biggl(\frac{g \lambda_V\,}
{M_W^2}\Biggr)^2\,
\frac{\Lambda^4}{128\,\pi^2}\,+\frac{3\,g}{16\,\pi^2}\,\frac
{g\,\lambda_V\,\Lambda^2}{M_W^2}\Biggr);
$$
That is, if $\lambda_V\,\ne\,0$
\be
1\,=\,\Biggl(\frac{\lambda_V}{\lambda_0}\Biggr)^2\,+\,
\frac{3\,g}
{\sqrt{2}\,\pi}\,\frac{\lambda_V}{\lambda_0}\,;
\label{compW1}   
\ee
and from~(\ref{compW1}) we have
\be
\lambda_V\,\simeq\,\lambda_0\biggl(\pm\,1\,-\,\frac{3\,g}   
{2\,\sqrt{2}\,\pi}\biggr)\,.\label{lamcor}
\ee
Value of $g$~(\ref{G11}) gives the correction term here to be 
about 20\%, what corresponds to data presented in Table 1.
We consider result~(\ref{lamcor}) as qualitative argument for
possibility of consistent description of data in the 
framework of our approach. Full study of this problem needs
special efforts. So for the moment one may only state, that 
value
\be
\lambda_V\,\simeq\,-\,0.034\,;\label{lamV}
\ee
which excellently fits $t$-quark mass, is possible in the 
approach
being considered here.
Sign of $\lambda_V$ is chosen due to experimental limitations
$\lambda_V\,=\,-\,0.037\,\pm\,0.030$~\cite{Osaka}. We see, 
that our
result~(\ref{lamV}) fits this restriction quite nicely.
We consider this result~(\ref{lamV}) as a prediction for 
triple gauge coupling.
   
We have no proof that the possibility being discussed here 
is the only one, which can improve situation with the 
normalization condition.
However all other possibilities, which were considered in 
the course of
performing the work, lead to wrong sign of corresponding 
contributions to
this condition. That is $t$-quark mass becomes even larger, 
than
$513\,GeV$. So we are inclined to consider this possibility, 
especially
prediction~(\ref{lamV}) as quite promising. In any case we 
have an
example of a variant, which has no contradiction with all 
what we know.

\section{Muon g-2}

Let us start with anomalous triple gauge coupling~(\ref{WWW}) 
and consider possible  additional interactions of composite 
Higgs with charged and neutral $W$-s 
\be
 f_{ch}\,H W_{\mu \nu}^* W_{\mu \nu};\qquad
f_0\,H W_{\mu \nu}^0 W_{\mu \nu}^0.\label{fWW}
\ee

Let us consider at first "zero" values of parameters $f_i$, 
which are defined by interactions already introduced above.

\begin{picture}(160,50)
\multiput(10,40)(2,0){10}%
{\circle*{1}}
\put(35,40){(W)}
\multiput(60,40)(0.5,0){40}%
{\circle*{1}}
\put(85,40){(H)}
\multiput(50,20)(0,0.5){20}%
{\circle*{1}}
\multiput(40,2.8)(2,0){10}%
{\circle*{1}}
\multiput(50,20)(-1,-1.72){16}%
{\circle*{1}}
\multiput(50,20)(1,-1.72){16}%
{\circle*{1}}
\put(50,20){\circle*{1}}
\put(40,2.8){\circle*{3}}
\put(60,2.8){\circle*{3}}
\end{picture}
\bigskip

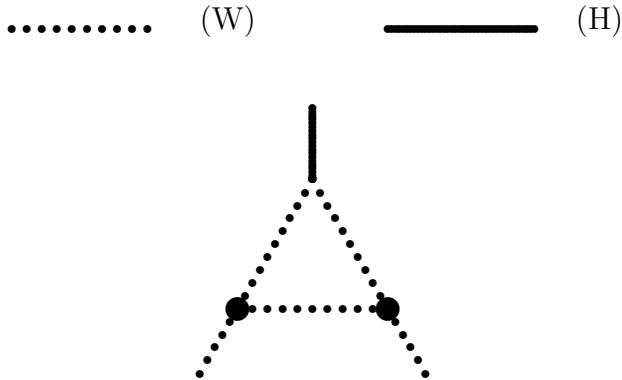
Figure 6. Diagram for zero approximation of $H W W$ 
interactions.\\
\\
  
From one-loop diagram Fig. 6 we have corresponding vertices
$$
V_i^0\,=\,f_i^0\,(g_{\mu \nu}(k\,q)- k_\nu q_\mu);
$$
where
\beq
& &f_0^0\,=\,\frac{g M_W}{8 \pi^2}\Biggl(\frac{g \lambda_V}
{M^2_W}\Biggr)^2\,\Lambda^2;\nonumber \\
& &f_{ch}^0\,=\,\frac{2\,g M_W}{8 \pi^2}\Biggl(\frac{g 
\lambda_V}
{M^2_W}\Biggr)^2\,\Lambda^2;
\label{zero}
\eeq
Here $k,\,q,\,\mu,\,\nu$ are respectfully momenta and 
indices of $W$-s. Other parameters are defined 
above. 
Diagrams Fig. 7 
with use of numbers~(\ref{zero})  
gives contribution to muon $(g-2)$ of order 
of magintude $\simeq 2\cdot 10^{-10}$, that is too small 
in comparison with experimental data~
\cite{muon} $(43 \pm 16)\cdot 10^{-10}$.

Let us consider one-loop equation in quasi-averages method. 
The corresponding compensation equation, presented at Fig. 8, 
reads
\be
f_i\,=\,f^0_i\,+\,\frac{\Lambda^2}{32\,\pi^2}\,f_i^3.
\label{comp}
\ee
\begin{picture}(160,50)
\put(60,40){\line(2,0){20}}
\put(85,40){($\mu$)}
\multiput(40,20)(0,2){5}%
{\circle*{1}}
\put(20,2.8){\line(2,0){40}}
\multiput(40,20)(-0.25,-0.43){40}%
{\circle*{1}}
\multiput(40,20)(1,-1.72){10}%
{\circle*{1}}
\put(40,20){\circle*{3}}
\put(30,2.8){\circle*{1}}
\put(50,2.8){\circle*{1}}
\put(70,10){+}
\multiput(100,20)(0,2){5}%
{\circle*{1}}
\put(80,2.8){\line(2,0){40}}
\multiput(100,20)(-1,-1.72){10}%
{\circle*{1}}
\multiput(100,20)(0.25,-0.43){40}%
{\circle*{1}}
\put(100,20){\circle*{3}}
\put(90,2.8){\circle*{1}}
\put(110,2.8){\circle*{1}}
\end{picture}
\bigskip

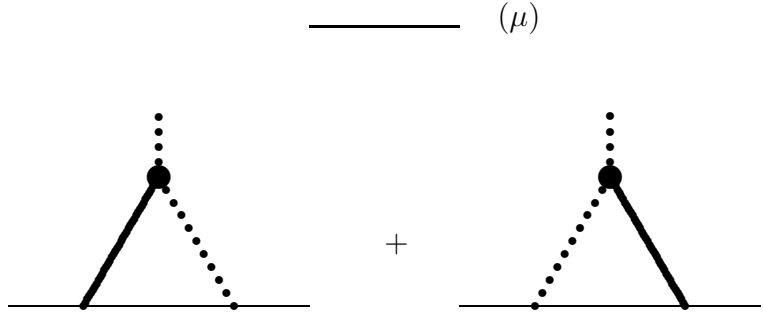
Figure 7. Diagrams for $H W^0 W^0$ contribution to 
the muon anomalous magnetic moment.\\
\\
\begin{picture}(160,40)

\multiput(10,20)(0,2){7}%
{\circle*{1}}
\multiput(10,20)(-0.43,-0.25){28}%
{\circle*{1}}
\multiput(10,20)(1.72,-1){7}%
{\circle*{1}}
\put(10,20){\circle*{3}}
\put(30,20){=}
\multiput(60,30)(0,2){5}%
{\circle*{1}}
\multiput(50,12.8)(2,0){10}%
{\circle*{1}}
\multiput(50,12.8)(-1,-1.72){6}%
{\circle*{1}}
\multiput(60,30)(-0.25,-0.43){40}%
{\circle*{1}}
\multiput(60,30)(1,-1.72){10}%
{\circle*{1}}
\multiput(70,12.8)(0.25,-0.43){20}%
{\circle*{1}}
\put(60,30){\circle*{3}}
\put(50,12.8){\circle*{3}}
\put(70,12.8){\circle*{3}}
\put(90,20){+}
\multiput(120,30)(0,2){5}%
{\circle*{1}}
\multiput(110,12.8)(2,0){10}%
{\circle*{1}}
\multiput(120,30)(-1,-1.72){10}%
{\circle*{1}}
\multiput(110,12.8)(-0.25,-0.43){20}%
{\circle*{1}}
\multiput(120,30)(0.25,-0.43){40}%
{\circle*{1}}
\multiput(130,12.8)(1,-1.72){6}%
{\circle*{1}}
\put(120,30){\circle*{3}}
\put(110,12.8){\circle*{3}}
\put(130,12.8){\circle*{3}}
\end{picture}
\bigskip

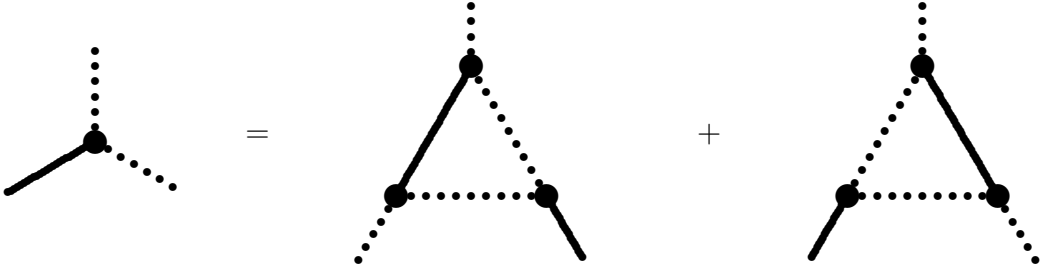
Figure 8. Diagram representation of equations for $f_i$ in 
the quasi-averages method.\\
\\
Here $f^0_i$ are quite small, so there are three 
solutions
\be
f_{i,1}\,=\,f^0_i;\quad 
f_{i,(2,3)}\,=\,\pm\,\frac{4 \sqrt{2}\,\pi}{\Lambda}.
\label{solution}
\ee
For solutions $f_{0,(2,3)}$ we have again from diagrams 
Fig. 7
\be
\Delta a\,=\,\pm\,\frac{-\,\sqrt{2}\,g\,m_\mu^2}
{16\,\pi\,\Lambda\,M_W}\,\Biggl(\,\ln\frac{\Lambda^2}{M_H^2}
-\frac{(1-4\sin^2\theta_W)M_Z^2}{M_H^2 - M_Z^2}\,
\ln{\frac{M_H^2}{M_Z^2}}\,\Biggr).
\label{da}
\ee
Substituting into~(\ref{da}) 
$$
g=0.653;\quad M_H=115\,GeV;\quad 
\Lambda\,=\,\frac{2 \sqrt{3}\,\pi\,M_H}{g\,\tan\theta_W}
\,=\,3506\,GeV;
$$
and usual values for $m_\mu,\,M_W$, we have
\be
\Delta a\,=\,\pm\,(-\,49.4\cdot 10^{-10}).\label{good}
\ee
The agreement is obvious (for sign minus 
in~(\ref{solution})). 

In other notations we have
\beq
& &\Delta a\,=\,\pm\,\frac{-\,\alpha\,m_\mu^2}
{4 \sqrt{6}\,\pi \sin\theta_W\,\cos\theta_W\,M_H\,M_W}\,
\Biggl(\,
\ln\frac{3\,\pi\,\cos^2\theta_W}{\alpha} - \nonumber\\
& &-\,\frac{(1-4\sin^2\theta_W)M_Z^2}{M_H^2 - M_Z^2}\,
\ln{\frac{M_H^2}{M_Z^2}}\,\Biggr);\label{formula}
\eeq
From here we may reformulate the result in other way. 
Let us take data~\cite{muon} and define $M_H$. We have 
for overall sign plus
\be
M_H\,=\,132^{+78}_{-37};\label{MH}
\ee

Sign in Eq.~(\ref{solution}) is an important 
point. It is defined by stability condition, which is 
already used in Section 3. Remind, that the larger is 
$|\lambda_V|$ the more stable is the solution and vice versa. 
Now, let us consider 
contributions to the compensation equation 
(\ref{compW}) for $\lambda_V$. According to 
diagrams containing vertices (\ref{fWW}) we have 
instead of (\ref{compW})
\be
\lambda_V\,=\,\lambda_V\,\frac{f_{ch}(f_{ch}+f_0)\,\Lambda^2}
{ 32 \,\pi^2}\,+\,\frac{\lambda_V}{128 \pi^2}\Biggl(
\frac{g \lambda_V \Lambda^2}{M_W^2}\Biggr)^2\,.
\label{fc0}
\ee

In view of~(\ref{solution}) we have $f_0 = \pm\,f_{ch}$ 
for nontrivial solutions for both $f_i$. 
For plus sign we have positively definite first term in 
the rhs. of~(\ref{fc0}). Substituting nontrivial 
solution~(\ref{solution}) we come to coefficient 2 
afore the $\lambda_V$ term in the rhs., 
and thus there 
is no nontrivial solution of the equation at all. So 
we have $\lambda_V\,=\,0$, that corresponds to 
unstable situation. For minus sign the contribution vanishes 
and one have to consider next approximations. The same is 
valid for trivial charged solution $f_{ch}=0$. For the last 
case the next terms are defined by contributions of 
one-loop diagrams Fig. 9, in which the upper line
 corresponds to neutral gauge boson. We would draw attention 
to the fact, that interaction vertex $H W^+ W^-$ is the 
usual one. Thus we have
\beq
& &\lambda_V\,=\,\lambda_V\, \frac{g M_W f_0}{48 \pi^2}\,
\ln\frac{\Lambda^2}{M_H^2}\,+\,
\lambda_V\,\Biggl(\frac{g\,\lambda_V}{M_W^2}\Biggr)\,
\frac{\Lambda^4}{128\,\pi^2}\,=\nn\\
& &=\,\pm\,\lambda_V\,\frac{\sqrt{2}\,g\,M_W}
{12\,\pi\,\Lambda}\,\ln\frac{\Lambda^2}{M_H^2}\,+\,
\lambda_V\,\Biggl(\frac{g\,\lambda_V}{M_W^2}\Biggr)\,
\frac{\Lambda^4}{128\,\pi^2}\,.\label{pm}
\eeq

This means, that just for minus sign of $f_0$ nontrivial 
solution for 
$|\lambda_V|$ becomes larger, and thus, according to 
discussion in Section 3 with the use of Table 1, the 
minimum becomes deeper. 

\begin{picture}(160,75)
\multiput(30,50)(0,2){7}%
{\circle*{1}}
\multiput(30,50)(-1.72,-1){7}%
{\circle*{1}}
\multiput(30,50)(1.72,-1){7}%
{\circle*{1}}
\put(30,50){\circle*{3}}
\put(50,50){=}
\multiput(80,60)(0,2){5}%
{\circle*{1}}
\multiput(70,42.8)(2,0){10}%
{\circle*{1}}
\multiput(80,60)(-1,-1.72){16}%
{\circle*{1}}
\multiput(80,60)(1,-1.72){16}%
{\circle*{1}}
\put(80,60){\circle*{3}}
\put(70,42.8){\circle*{3}}
\put(90,42.8){\circle*{3}}
\put(110,50){+}
\multiput(20,20)(0,2){5}%
{\circle*{1}}
\multiput(10,2.8)(2,0){10}%
{\circle*{1}}
\multiput(10,2.8)(-1,-1.72){6}%
{\circle*{1}}
\multiput(20,20)(-0.25,-0.43){40}%
{\circle*{1}}
\multiput(20,20)(1,-1.72){10}%
{\circle*{1}}
\multiput(30,2.8)(1,-1.72){6}%
{\circle*{1}}
\put(20,20){\circle*{3}}
\put(30,2.8){\circle*{3}}
\put(50,10){+}
\multiput(80,20)(0,2){5}%
{\circle*{1}}
\multiput(70,2.8)(2,0){10}%
{\circle*{1}}
\multiput(80,20)(-1,-1.72){10}%
{\circle*{1}}
\multiput(70,2.8)(-1,-1.72){6}%
{\circle*{1}}
\multiput(80,20)(0.25,-0.43){40}%
{\circle*{1}}
\multiput(90,2.8)(1,-1.72){6}%
{\circle*{1}}
\put(80,20){\circle*{3}}
\put(70,2.8){\circle*{3}}
\end{picture}
\bigskip

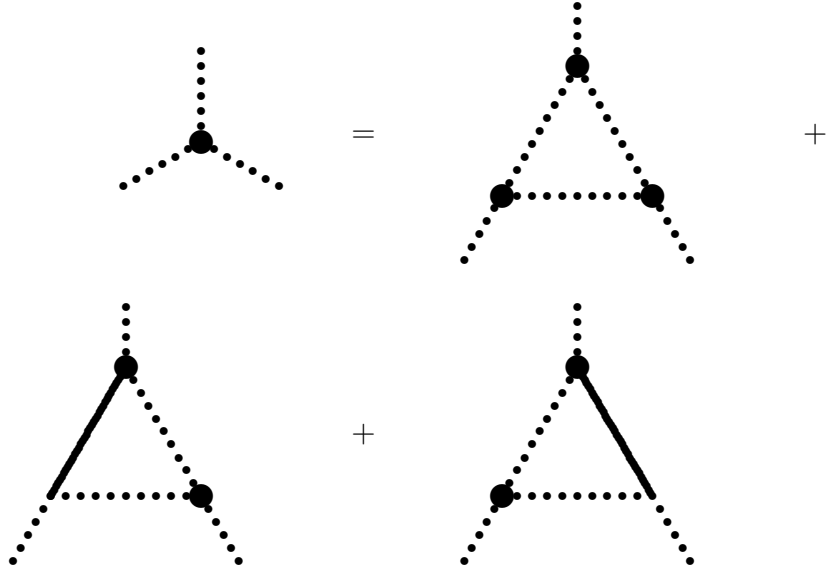
Figure 9. contributions of $f_0$ to equation for 
$\lambda_V$.\\
\\
We easily see, that for possibility 
$f_{ch}\,=\,-\,f_0$ the contribution becomes smaller in 
absolute value because due to opposite signs of $f_0$ and 
$f_{ch}$ their contributions almost cancel. 
Thus this case corresponds to the less 
deep minimum. We come to the conclusion, that the most 
stable solution corresponds to
\be
f_{ch} \,=\,0;\qquad f_0\,=\,-\,\frac{4\sqrt{2} \pi}
{\Lambda};\qquad \Delta a\,=\,49.4\cdot 10^{-10}.
\label{final}
\ee
So we obtain the result, that 
our approach chooses minus sign of~(\ref{solution}) and this 
means plus sign in Eqs.~(\ref{da}, \ref{formula}) in full 
correspondence with experimental indications~\cite{muon}.

Finally, we have the following vertices  
of Higgs particle $H$ with gauge bosons $Z,\,A$
\beq
& &H\,A\,A\,:\quad -\,\sin^2\theta_W\,
\frac{4 \sqrt{2}\,\pi}{\Lambda}\,(g_{\mu \nu}(k\,q)- 
k_\nu q_\mu);\nonumber\\
& &H\,Z\,A\,:\quad -\,\sin\theta_W\,\cos\theta_W\,\frac
{4 \sqrt{2}\,\pi}{\Lambda}\,(g_{\mu \nu}(k\,q)- 
k_\nu q_\mu);\label{vert}\\
& &H\,Z\,Z\,:\quad \frac{g\,M_W}{\cos^2\theta_W}\,
g_{\mu \nu}\,-\,\cos^2\theta_W\,\frac{4 \sqrt{2}\,\pi}
{\Lambda}\,(g_{\mu \nu}(k\,q)- k_\nu q_\mu);
\nonumber
\eeq
where $k,\,q,\,\mu,\,\nu$ are respectfully momenta and 
indices of gauge bosons. 

Now, interactions~(\ref{vert}) radically change branching 
ratios of $H$ decays. For $M_H\,=\,115\,GeV$ we present 
results at Table 2.

Interactions~(\ref{vert}) lead also to a change in 
predictions for Higgs production. For $e^+e^-$ 
reactions we present results of calculations for three 
energies: 189 GeV, 207 GeV (maximal luminosity at LEP2) 
and 250 GeV (to demonstrate future possibilities). 
In calculations ISR is taken into account. The results 
are presented in Table 3.
\newpage
\begin{center}
Table 2.\\
Branching ratios for 115 GeV Higgs decays \\
in the present model (new)\\ 
and in Standard Model (SM).

\bigskip

\begin{tabular}{|l|c|c|}\hline
decay channel & B.R. new \% & B.R. SM \% \\ \hline
$H \to \gamma\,\gamma$ & 62.8 & 0.3 \\ \hline
$H \to \gamma\,Z$ & 22.3 & 0.1 \\ \hline
$H \to b \bar b$ & 11.7 &  79.9\\ \hline
$H \to \tau \bar \tau$ & 1.6 & 9.3\\ \hline
$H \to c \bar c$  & 0.4 & 2.5\\ \hline
$H \to g\,g$ & 0.1 & 0.9 \\ \hline
$H \to W^* W^*$ & 0.9 & 6.0\\ \hline
$H \to Z^* Z^*$ & 0.2 & 1.0\\ \hline
\end{tabular}
\end{center}
\bigskip

\begin{center}
Table 3.\\
Cross-sections (pb) of Higgs production reactions \\
for three different energies 
in the present \\
model (new) and in Standard Model (SM).

\bigskip

\begin{tabular}{|l|c|c|c|}\hline
reaction & 189 GeV & 207 GeV & 250 GeV\\ \hline
$e^+e^-\to H\,Z$ new& 0.000 & 0.300 &  1.997 \\ \hline
$e^+e^-\to H\,Z$ SM& 0.000 & 0.057 &  0.273\\ \hline
$e^+e^-\to H\,\gamma$ new & 0.124 & 0.157  & 0.217 \\ \hline
$e^+e^-\to H\,\gamma$ SM & 0.000 & 0.000  & 0.000 \\ \hline
$e^+e^-\to e^+e^-\,H$ new& 0.059 & 0.080 & 0.180 \\ \hline
$e^+e^-\to e^+e^-\,H$ SM& 0.000 & 0.002 & 0.010\\ \hline
$e^+e^-\to H + X$ (total) new & 0.18 & 0.54 & 2.39 \\ \hline
$e^+e^-\to H + X$ (total) SM & 0.00 & 0.06 & 0.28\\ \hline
\end{tabular}
\end{center}
\bigskip

We see from Tables 2 and 3, that for real search~\cite{New} 
of Higgs in the channel $H \to b \bar b,\;\sigma\cdot B.R$ 
are $0.035\,pb$ (new) and $0.046\,pb$ (SM) that means 
almost the same. By the way, the 25\% decrease of the 
new result in comparison with SM prediction may serve 
for an explanation of rather indefinite evidence~\cite{New} 
for $115\,GeV$ Higgs. 
However, in the channel $H \to \gamma\, \gamma$ there is 
very considerable effect. For $\sqrt{s}=189\, GeV$, 
where there are 
data~\cite{L3},~\cite{L32} $\sigma(e^+e^-\to H \gamma) 
\cdot BR(H\to 2\gamma) 
< 0.122\,pb$ for $M_H \simeq 110\,GeV$  
to be compared 
with  our prediction $\sigma(H\,\gamma)\cdot BR(2 \gamma) = 
0.078\,pb$. So, there is no contradiction with data.
For $\sqrt{s}= 207 \,GeV$ we predict $\sigma(H\,
\gamma)\cdot BR(2 \gamma) = 0.099\,pb$ that means e.g. 
around 14  events in the already 
collected L3 data. So it seems advisable 
to develop these data 
and look for channel $e^+e^- \to 3\,\gamma$, two of 
the photons being due to decay $H \to  \gamma\,\gamma$.

We calculate also effect of the model for Higgs production 
at FNAL. Corresponding cross-sections are presented at 
Table 4. We see a considerable enhancement of the production 
process in our model. Thus a check of the model at FNAL might 
be possible provided integral luminosity being few tens of 
$pb^{-1}$.

\begin{center}
Table 4.\\
Cross-sections of Higgs production in $p \bar p$ 
reactions \\
at $\sqrt{s}= 2000\,GeV$ in the present model (new) \\
and in Standard Model (SM).

\bigskip

\begin{tabular}{|l|c|c|}\hline
reaction & $\sigma\,pb$ new & $\sigma\,pb$ SM \\ \hline 
$p \bar p \to H\,Z + X$ & 0.97 & 0.10 \\ \hline
$p \bar p \to H\,W^\pm + X$ & 0.17 & 0.17 \\ \hline
$p \bar p \to H\,\gamma + X$ & 0.17 & 0.00 \\ \hline
$p \bar p\to p \bar p\,H + X$ & 0.35 & 0.08\\ \hline
$p \bar p \to H + X$ (total)& 1.66 & 0.35\\ \hline
\end{tabular}
\end{center}

Note, that the enhancement of production 
processes is connected with the negative sign in main 
quantity $f_0$~(\ref{solution}). For the positive sign 
we have just the opposite situation of suppression 
of $H$ production. In this case one cannot even interpret 
the indications for 115 GeV Higgs~\cite{New}. 
So once more we would emphasize the 
importance of this sign definition, which in our 
model follows from the basic property of stability 
of a scalar condensation.

\section{Discussion}

Thus as a result of our study of the model we now come to 
the theory,  
which describe Standard electroweak interaction of:\\
1. gauge vector bosons sector with input of constants $g$ 
and   $\theta_W$;\\
2. heavy quark doublet $t,\,b$ with  the $t$-quark mass
$174\,\pm\, 5\,GeV$ and the $b$ mass zero;\\
3. necessary composite Higgs sector with mass of surviving 
neutral scalar particle $ M_H\,=\,117\, \pm \,7\,GeV$.

In addition to these standard interactions there are also 
extra effective interactions, which act in momentum 
regions limited by effective cut-off 
$\Lambda\,\simeq\,3500\,GeV$.
We calculate the last  number e.g. from 
relation~(\ref{m0n}).   
Three parameters ($g,\,\theta_W,\,\Lambda$) are the only
fundamental input of the model. As a technical input we 
consider
triple gauge constant $\lambda_V$, which in range of 20\% 
agrees  with the calculated value~(\ref{lV}), but its 
precise value is
defined by the next approximations. We emphasize, that the 
presence of such interaction with constant~(\ref{lamV})
in the framework of the model is necessary.
We consider result~(\ref{lamV}) as very important prediction 
of the
model. Of course, the most important for the model is the 
prediction, that hints for $115\,GeV$ Higgs~\cite{New} 
have to be confirmed. We also would emphasize the importance 
of prediction for new contribution to the muon anomalous 
magnetic moment $\Delta a$~(\ref{final}) , 
which agrees with experimental data. Of course, the 
experimental status of the effect is not yet decisive and 
needs further clarification. Arguments are expressed 
repeatedly, that uncertainties in hadronic contributions 
to $g-2$ do not allow to insist on discrepancy with SM (see, 
e.g.~\cite{APiv}). However, the result being obtained 
here does not contradict to any calculation of the effect, 
including that of~\cite{APiv}. 

It is worthwhile here to comment the problem of accuracy of 
the approach. Our approximations are the following.\\
1. One-loop diagrams. We estimate accuracy to be around
$\bar X_1\,\simeq\,0.08$. \\
2. Corrections being of order of magnitude $g/\pi$ give
uncertainty around 0.2.\\
3. The leading order of $1/N_c$ expansion. Usually precision 
of $1/N_c$ expansion is estimated to be $1/N_c^2\,\simeq\, 
0.11$.\\
4. We keep only logarithmic terms. Uncertainty is estimated 
to be $1/\log(\Lambda^2/m_t^2)\,\simeq\,0.16$.

Thus the combined accuracy, provided items 1 -- 4 being 
independent, is estimated to be around 29\%. As a matter 
of fact, all uncertainties,
which we have encountered above fit into this range.

For the moment we have only one quark doublet. Remind, that 
now we understand why the $t$-quark is heavy and the 
$b$-quark is almost
massless. We consider heavy $t,\,b$ doublet as the most
fundamental one in the sense, that it defines the main
part of heavy particles masses. Indeed, composite Higgs 
scalars are just
consisting of these quarks. We may immediately introduce 
all other quarks
and leptons with zero mass completely in the line
prescribed by the Standard Model. So at this stage of study of
the model we deal with massive $W,\,Z,\,t,\,H$ and massless 
all other  particles. We may expect,that masses of quarks 
and leptons will
appear in subsequent approximations. We have to expect their 
value to be less than 29\% of our heavy masses, that is 
$m_i\,\leq\,23\,GeV$. All
the masses known satisfy this restriction. We could expect 
also, that in
subsequent approximations Higgs scalars are not composed 
of only heavy
quarks, but have admixture of lighter quarks and leptons.

Additional contact interactions~(\ref{nset}) include
four-fermion interactions of heavy quarks $t,\,b$. So, 
there is
practically no experimental limitations on the corresponding
coupling constant. Limitations exist only for contact 
interactions
of light quarks. Limitations for triple gauge 
interaction~(\ref{WWW})
were mentioned above and were shown to agree with predicted
value~(\ref{lamV}). 

Experimental consequences of interactions~(\ref{fWW}) are 
discussed above and shown to have no contradiction with the 
present data. 

To conclude the author would state, that the variant being 
studied
presents a possibility to  formulate realistic electroweak 
interaction
without elementary scalars, which are substituted by 
composite effective scalar fields. The dynamics of the 
variant is defined
in the framework of the Bogolyubov quasi-averages method. 
Consequences of the approach contains no contradiction with 
the existing data and agree with two would-be effects: 
$115\,GeV$ Higgs indications and evidence for deviation 
from SM prediction in muon $g-2$. Definite
predictions make checks of the variant to be quite 
straightforward.
\begin{flushleft}
{\Large \bf Appendix}
\end{flushleft}
\appendix

Let us consider a theory with combination of four-fermion
interactions~(\ref{psi}). We take the simplest nontrivial 
term in the Bethe-Salpeter equation for connected 
four-fermion amplitude,
corresponding to Lorentz structure
$\bar \psi_L \gamma_\rho \psi_L\,\bar \psi_R \gamma_\rho 
\psi_R\,=\,
-\,2\,\bar \psi_L \, \psi_R\,\bar \psi_R \, \psi_L\,$,
which is presented at Fig. 10.
 
The kernel for this equation
corresponds to simple loop, which gives the following 
expression
\be
K(p,q)\,=\,\imath\,\frac{3\,x\,(y+z)}{16\,\pi^2}\,(p-q)^2\,
 \log\,(p-q)^2\,+\,const\,+\,{\cal O}\,((p-q)^2).
\label{kernel}
\ee
We are interested in momentum dependence, so we
consider just the logarithmic term in the kernel, i.e. the 
first term
n Eq.~(\ref{kernel}), which we denote $K_{log}(p,q)$. We shall
see below, that terms containing $const$ and $const\,(p-q)^2$
give zero contribution due to boundary conditions.
Equation for four-fermion amplitude F(p), corresponding to 
Fig. 10,
looks like (after applying of Wick rotation)

\begin{picture}(160,30)
{\thicklines
\put(30,10){\line(-1,1){10}}
\put(30,10.5){\line(1,1){10}}}
\put(30,10){\circle*{3}}
\put(30,10){\line(-1,-1){10}}
\put(30,9.5){\line(1,-1){10}}
\put(60,10){=}
{\thicklines
\put(100,10){\line(-2,1){30}}
\put(100,10){\line(-2,-1){30}}
\put(80,10){\oval(5,20)}
\put(80,20){\oval(2,4)}
\put(80,0){\oval(2,4)}
\put(100,10){\line(1,1){10}}
\put(100,10){\line(1,-1){10}}
\put(100,10){\circle*{3}}}
\end{picture}
\bigskip

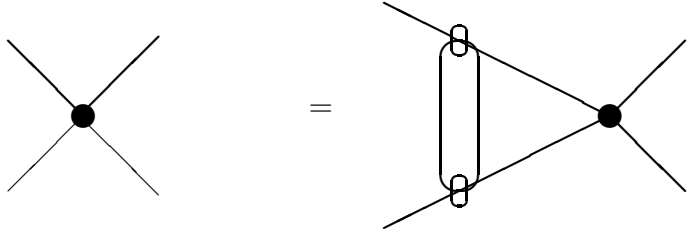
Figure 10. Bethe-Salpeter equation in quasi-ladder 
approximation.\\
\\
\beq
& &F(p)\,=\,-\,\imath \int\,\frac{K_{log}(p,q)\,F(q)}
{q^2}\,d^4q\,\,=
\,\beta \int\frac{(p-q)^2\,\log(p-q)^2\,F(q)}{q^2}\,d^4q\,;
\label{eqf}\\
& &\beta\,=\,\frac{3\,x\,(y+z)}{16\,\pi^2}\,.\nn
\eeq
Equations of such type can be studied in the same way as more
simple equations with kernels being proportional to 
$((p-q)^2)^{-1}$.
The method to solve the latter ones was known a long time
ago~\cite{AF}. For logarithmic case basic angular integrals 
are the following
\beq
& &\int d\Omega_4\,\log(p-q)^2\,=\,\pi^2\biggl(\theta(x-y) 
\biggl(\frac{y}{x}+2 \log\,x\biggr)\,+\,\theta(y-x)
\biggl(\frac{x}{y}+2 \log\,y\biggr)\biggr)\,;\nn\\
& &\int d\Omega_4\,(p\,q)\,\log(p-q)^2\,=\,\frac{\pi^2}{3}
\biggl(\theta(x-y)
\biggl(\frac{y^2}{x} - 3\,y\biggr)\,+\,\theta(y-x)
\biggl(\frac{x^2}{y} - 3\,x\biggr)\biggr)\,;\label{log}\\
& &x\,=\,p^2,\qquad y\,=\,q^2\,.\nn
\eeq
Integrals~(\ref{log}) are sufficient for angular 
integrations in equation~(\ref{eqf}). So we have
\beq
& &F(x)\,=\,\frac{\beta}{32\,\pi^2}\biggl(\frac{1}{3\,x}
\int_0^x
y^2\,F(y)\,dy\,+\,3\int_0^x y\,F(y)\,dy\,+\,2\,\log\,x\,
\int_0^x
y\,F(y)\,dy\,+\nn\\
& &+\,2\,x\,\log\,x\,\int_0^x F(y)\,dy\,+\,2\,\int_x^\infty
y\,\log\,y\,F(y)\,dy\,+\,x \int_x^\infty (3 + 2 \log\,y)\,
F(y)\,
dy\,+\nn\\
& &+\,\frac{x^2}{3}\int_x^\infty \frac{F(y)}{y}\,dy\biggr)
\,.\label{eq1}
\eeq
Applying differentiations in proper order, we obtain 
differential equation
\be
\frac{d^3}{dx^3}\,\Bigl(x^2\,\frac{d^3}{dx^3}\,(x\,F(x))
\Bigr)\,=
\,-\,\frac{\beta}{8\,\pi^2}\,\frac{F(x)}{x}\,.\label{dif}
\ee
Eq.~(\ref{dif}) is easily transformed to the canonical form 
of Meijer equation~\cite{BE}
\beq
& &\biggl(y\,\frac{d}{dy}+\frac{1}{2}\biggr)\,
\biggl(y\,\frac{d}{dy}\biggr)\,
\biggl(y\,\frac{d}{dy}\biggr)\,
\biggl(y\,\frac{d}{dy}-\frac{1}{2}\biggr)\,
\biggl(y\,\frac{d}{dy}-\frac{1}{2}\biggr)\,
\biggl(y\,\frac{d}{dy}-1\biggr)\,F(y)\,+\nn\\
& &+\,y\,F(y)\,=\,0\,;
\qquad y\,=\,\frac{\beta}{512\,\pi^2}\,x^2\,.\label{Meyer}
\eeq
Equation~(\ref{dif}) (or~(\ref{Meyer})) is equivalent to 
integral
equation~(\ref{eq1}) provided proper boundary conditions be
fulfilled. The first one is connected with convergence of 
integrals
at infinity. The most intricate conditions are connected with
behaviour at $x\to 0$.
From Eq.~(\ref{Meyer}) we see, that at zero there are the
following independent asymptotics:
\be
\frac{a_{-1}}{x},\;a_{0\,l}\,\log\,x,\;a_0,\;a_{1\,l}\,
x\log\,x,\;
a_1\,x,\;a_2\,x^2\,.\label{x0}
\ee
To obtain conditions for $a_i$ one has to substitute into 
Eq.~(\ref{eq1})
$$
F(y)\,=\,-\,\frac{8\,\pi^2}{\beta}\,x\,\frac{d^3}{dx^3}\,
\Bigl(x^2\,
\frac{d^3}{dx^3}\,(x\,F(x))\Bigr)\,;
$$
and perform integrations by parts. Thus we come to conditions
\be
a_{-1}\,=\,a_{0\,l}\,=\,a_{1\,l}\,=\,0\,.\label{cond}
\ee
These results, in particular, lead to zero values of integrals
$$
\int_0^\infty\,F(y)\,dy\,=\,\int_0^\infty\,y\,F(y)\,dy\,=
\,0\,;
$$
that guarantees the absence of contributions of constant 
terms in the initial equation, which we have mentioned above.

According to properties of Meijer functions~\cite{BE} we 
have unique solution
of Eq.~(\ref{Meyer}) with the boundary conditions~(\ref{cond})
\be
F(x)\,=\,\frac{\sqrt{\pi}}{2}\,G_{0\,6}^{3\,0}\Bigl(\,y\,|\,  
0,\,\frac{1}{2},\,1\,;\,- \frac{1}{2},\,0,\,\frac{1}{2}\,
\Bigr)\,;
\qquad y\,=\,\frac{\beta}{512\,\pi^2}\,x^2\,.
\label{sol}
\ee
This function has all qualities of a form-factor. It is 
equal to unity
at the origin and decreases with oscillations at infinity. 
Effective
cut-off is  estimated from~(\ref{sol})
\be
\Lambda^2\,=\,\frac{16\,\sqrt{2}\,\pi}{\sqrt{\beta}}\,.
\ee

Thus we demonstrate how the effective cut-off arises. 
Of course, it
is done in the quasi-ladder approximation, but we think, that
an account of non-linearities changes numbers, but does not 
change
the situation qualitatively. In the present work we will 
not go
further and will not try to connect these considerations with
realistic numbers.
   

\end{document}